\documentstyle[epsfig]{article}
\date{}
\def\Journal#1#2#3#4{{#1} {\bf #2}, #3 (#4)}

\def\NPB{{\em Nucl. Phys.} B}
\def\NPBP{{\em Nucl. Phys.} B (Proc. Suppl.)}
\def\NPA{{\em Nucl. Phys.} A}
\def\PLB{{\em Phys. Lett.}  B}
\def\PRL{\em Phys. Rev. Lett.}
\def\PRD{{\em Phys. Rev.} D}

\def\APJ{\em ApJ}
\def\EPL{\em Europhys. Lett.}

\def\ra{\rightarrow}
\def\be{\begin{equation}}
\def\ee{\end{equation}}
\def\bea{\begin{eqnarray}}
\def\eea{\end{eqnarray}}
\newcommand{\bb}{double beta decay }
\newcommand{\obb}{0\mbox{$\nu\beta\beta$ decay} }

\newcommand{\mas}{Majorana neutrinos }
\newcommand{\n}{neutrino }
\newcommand{\ns}{neutrinos }
\newcommand{\sit}{\mbox{$sin \theta$}}
\newcommand{\siqt}{\mbox{$sin^2 \theta$}}
\newcommand{\cost}{\mbox{$cos \theta$}}
\newcommand{\coqt}{\mbox{$cos^2 \theta$}}
\newcommand{\ema}{\mbox{$<m_{\nu_e}>$}}
\newcommand{\me}{\mbox{$m_{\nu_e}$}}
\newcommand{\mm}{\mbox{$m_{\nu_\mu}$}}
\newcommand{\mt}{\mbox{$m_{\nu_\tau}$}}
\newcommand{\ger}{\mbox{$^{76}Ge$ }}
\newcommand{\mo}{\mbox{$^{100}Mo$}}
\newcommand{\nd}{\mbox{$^{150}Nd$} }
\newcommand{\ca}{\mbox{$^{48}Ca$ }}
\newcommand{\teha}{\mbox{$^{128}Te$ }}

\newcommand{\xehs}{\mbox{$^{136}Xe$ }}
\title{MeV neutrinos in double beta decay}
\begin{document}
\author{K. ZUBER}
\maketitle
\begin{center}
{\it Lehrstuhl f\"ur Exp. Physik IV, Universit\"at Dortmund, 
Otto-Hahn-Str. 4,\\ 44221 Dortmund,
Germany}
\end{center}
\begin{abstract}
The effect of \mas in the MeV mass range on the
\bb of various isotopes is studied on pure
phenomenological arguments. By using only experimental half life 
data, limits on the mixing parameter $U_{eh}^2$ of the order 10$^{-7}$ can be derived.
Also the possible achievements of upcoming
experiments and some consequences are outlined.
\end{abstract}
One of the major challenges of modern physics is the still open question
of a nonzero neutrino mass. A massive \n could have important consequences
for astrophysics and cosmology, i.e. massive \ns are good dark matter 
candidates
and massive \ns are the preferred solution of the solar neutrino problem via the
MSW-effect. From the particle physics point of view massive \ns would require
a modification of the successful standard model and would open one of the most
promising ways for testing models beyond the standard model (GUT-theories).
 For the physical potential of massive \ns see \cite{books}.\\
Direct measurements of \n masses result at present in limits for the 
three \ns of
\bea
\me & < & 4.35 \quad eV \quad \mbox{(out of tritium beta decays \cite{trit})},\\ 
\mm & < & 160 \quad keV \quad \mbox{(out of pion-decays \cite{pion})}\\
\mt & < & 24 \quad MeV \quad \mbox{(out of tau-decays \cite{aletau})}
\eea
As can be seen MeV $\tau$-\ns are not ruled out at present. In recent 
times there is a growing interest in models with MeV $\tau$-\ns \cite{mike}. 
To be cosmological acceptable such \ns must be unstable because otherwise they
will overclose the
universe. On the other hand bounds on the number
of \n flavours coming out of big bang nucleosynthesis have been relaxed
recently, now allowing a value between about 2.2 \cite{hata} and 3.9 \cite{copi}.
This opens space for a MeV-mass of $\nu_{\tau}$.\\ 
In this paper the effects of a MeV Majorana-neutrino in \bb are 
investigated. For a discussion of heavy sterile
\ns in \bb see \cite{cliff}.
The analysis follows partly 
that of \cite{halp,leung}. Neutrinoless \bb (\obb) of a nucleus (A,Z)
\be
(A,Z) \ra (A,Z+2) + 2 e^-
\ee
violates lepton number by 2 units and is only
possible if \ns are Majorana particles (see \cite{books}). There are 
about 35 possible \bb emitters,
about 10 of them have experimental obtained half life limits of larger 
than 10$^{20}$y.
At present the best limit results from
the Heidelberg-Moscow collaboration studying the decay of \ger \cite{kk}
\be
T_{1/2}^{0\nu} > 7.4 \cdot 10^{24} a \ra \ema < 0.56 \quad eV \quad (90\% 
\quad CL) \ee 
The measured quantity (neglecting right handed weak currents), called 
effective Majorana mass $\ema$, is given 
in the case of light \ns ($m_{\nu} <$ 1 MeV) by
\be
\label{for5} \ema = \mid \sum_{i=1}^N U^2_{ei} m_i \mid
\ee
where $m_i$ characterizes the N mass eigenstates and $U_{ei}$ the mixing matrix
elements. Things change in case of heavy \ns ($m_{\nu} >$ 1 MeV). By calculating the
nuclear matrix elements for \bb involving 
MeV \ns the \n mass in the neutrino propagator can no
longer be neglected with respect to the \n momentum. 
For a detailed discussion on the matrix element calculations see \cite{muto}.
This results in a change
of the radial shape of the used \n potential H(r) from
\be
H(r) \propto \frac{1}{r} \quad \mbox{light \ns} \ra H(r) \propto 
\frac{exp(-m_hr)}{r} \quad \mbox{heavy \ns}
\ee

This changes can be accomodated for by introducing an additional factor 
$F (m_h,A)$
in eq. (\ref{for5}), which depends on the mass of the heavy \n $m_h$ and 
on the atomic number A of the nucleus. Eq. (\ref{for5}) is modified to
\be
\ema = \mid \sum_{i=1,light}^N U^2_{ei} m_i + \sum_{h=1,heavy}^M F (m_h,A) 
U^2_{eh} m_h  \mid
\ee
Assuming one heavy neutrino with $m_h = m_2$ the function $F (m_h,A)$ is given
by
\be
F (m_2,A) = <1/r>^{-1} <exp(-m_2r)/r>
\ee
r corresponds to the distance of the two nucleons in the nucleus undergoing 
\obb.
The average is with respect to the two nucleon correlation function appropriate
for the nucleus. Using a correlation function containing a hard core repulsion
characterized by a hard core radius $r_c$ of 0.5 fm between the two nucleons
\be
\rho(r) = \frac{1}{4/3 \pi [(2R)^3 - r_c^3]} \theta(r-r_c) \theta(2R-r)
\ee
it follows 
\be
\label{fh}F (m_2,A) = 0.5 \frac{1}{(m_2 R)^2} [(1+m_2 r_c)exp(-m_2r_c) - (1 
+ 2m_2R) exp(-2m_2R)]
\ee
For all nuclei of interest (A $\ge$ 48) the nuclear radius R is much 
larger than
the hard core radius (R $\gg r_c$). Therefore $F (m_2,A)$ varies with the nuclear
radius as R$^{-2}$ or using the relation R $\simeq 1.2 A^{1/3}$ fm it results in a 
dependence of A$^{-2/3}$.\\
Consider now the simple case of an electron coupled via the standard weak charged current to two massive \mas $\chi_{1,2}$ under the assumption of CP-conservation:
\be
\label{fo11} \nu_e = \chi_1 \cost + \chi_2 \sit 
\ee
where the fields $\chi_{1,2}$ satisfy the Majorana condition (C is the charge conjugation
matrix):
\be
\chi_{1,2} = \eta_{1,2} C \bar \chi_{1,2} \quad \eta_i = \pm 1
\ee
The corresponding $\ema$ is then given as
\be
\ema = \mid m_1 \coqt  + \eta_1 \eta_2 m_2 \siqt \mid 
\ee
in the case of only light \ns or for one light and a MeV-\n as
\be
\label{mh}\ema = \mid m_1 \coqt  + F(m_2,A) \eta_1 \eta_2 m_2 \siqt \mid 
\ee
In case of small mixing angles $\theta$ the $\beta$-decay experiments
measure $m_1$. For \bb if $\theta \neq 0$ and $\eta_1 \eta_2$ =
-1 something like destructive interference can occur in $\ema$ and the 
value measured can be smaller than the one from the tritium experiments.
\begin{figure}[hhh]
\begin{center}
\begin{tabular}{cc}
\epsfig{file=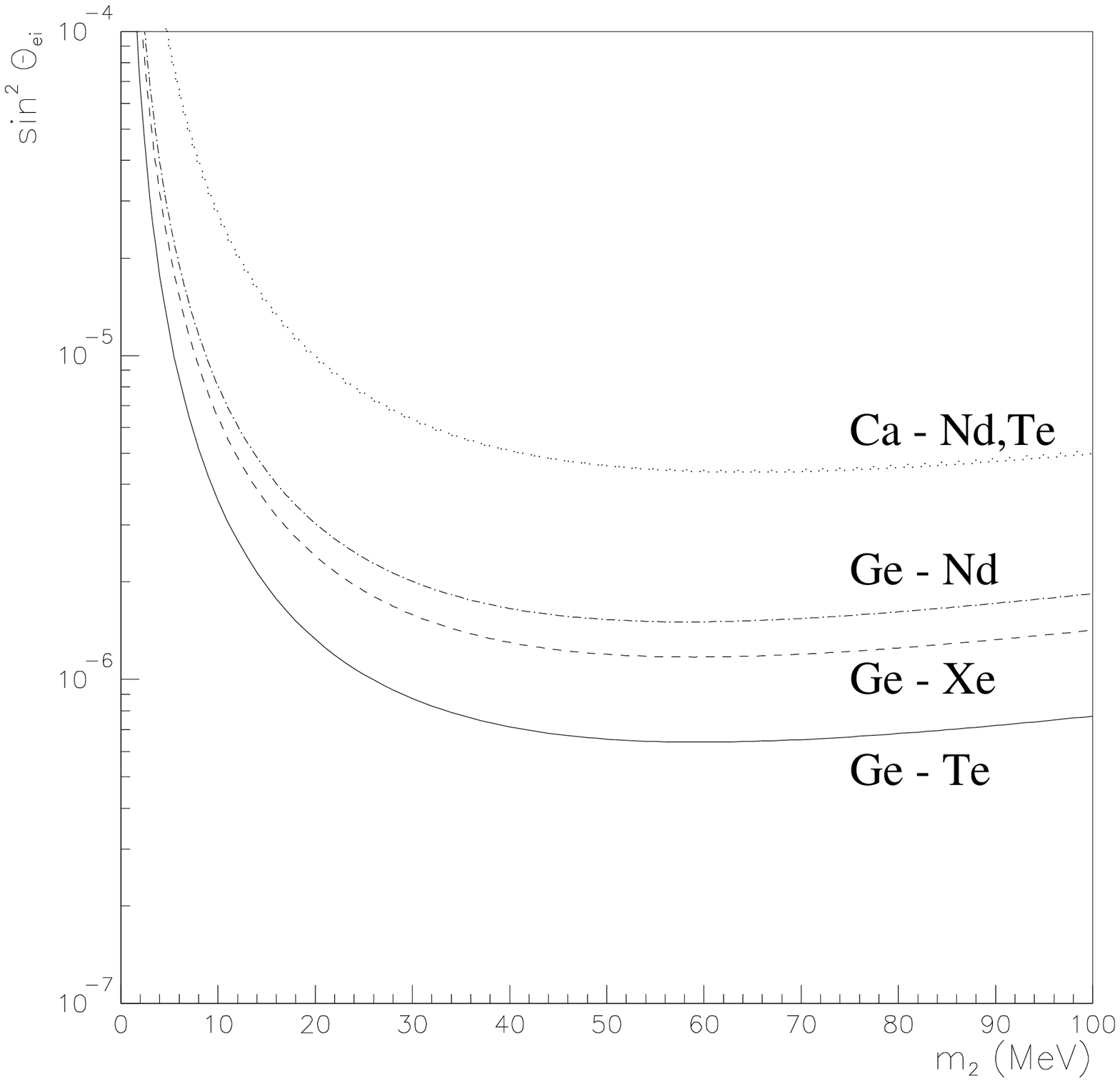,width=8cm,height=8cm}
& \epsfig{file=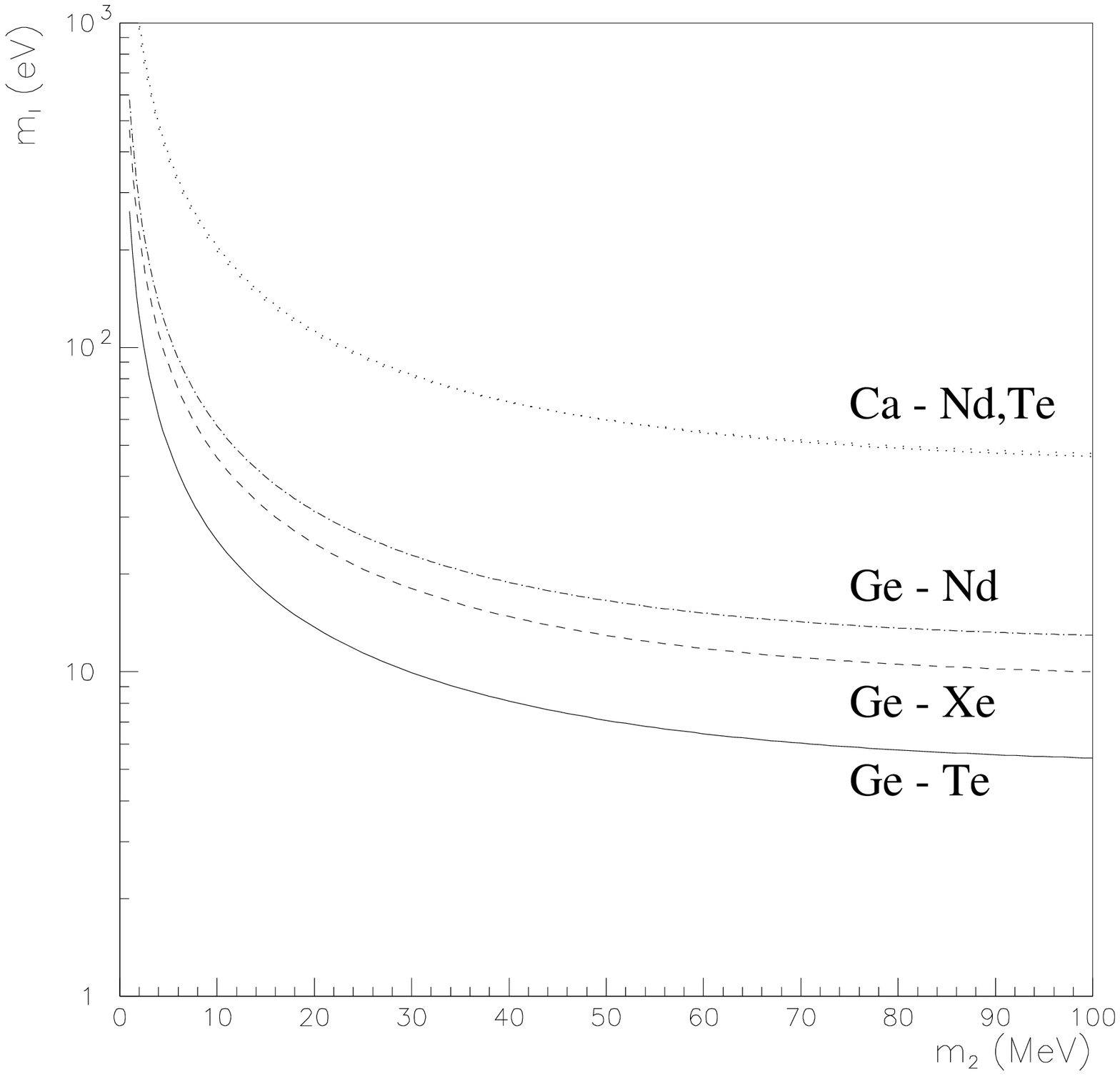,width=8cm,height=8cm}
\end{tabular}
\end{center}
\caption{\label{lego} \it left: Limit on the mixing angle sin$^2 
\theta_{ei}$ as a
function
 of the heavy \n mass $m_2$ in the region of 1-100 MeV for five 
combinations
 of isotopes. At present the best limit is given by the Ge-Te pair. 
The
combinations of the Ca-Nd and Ca-Te result
are nearly identical. right: Upper limit of the Majorana mass of
 a light neutrino $m_1$ in the case of interference with one heavy 
neutrino.
 The Ge-Te pair results in the case of $m_2 \approx 100 MeV$ in a 
limit
 for $m_1$ of about 6 eV.}
\end{figure}\\
Using eq.(\ref{fh}) and (\ref{mh}) it can be seen that $\ema$ gets an 
A-dependence
because of $F(m_2,A)$ making it worthwile to look into experimental
results of different \bb emitters. In Tab.1 a comparison of some
\bb emitters as well as present limits on the half life $T_{1/2}^{0\nu}$ 
and the effective \mas \ema  are shown.
\begin{center}
\begin{tabular}{|c|c|c|c|c|c|c|}
\hline
Isotope & $\ca$ & $\ger$ & $\mo$ & $ \teha$ & $\xehs$ & $\nd$
\\
\hline
$T_{1/2}^{0\nu}$ present & 9.5 E21& 7.4 E24 & 4.4 E22 & 7.7E24 & 4E23 & 
2.1E21\\ \hline
$\ema$ & 12.8& 0.56 & 5.4 & 1.0 & 2.4 & 4.0 \\
\hline
$T_{1/2}^{0\nu}$ future & 1 E23 & 1.5 E25 & 1E25 & 7.7 E24 & 3.6 E25 & 
1E23 \\ \hline
\end{tabular}
\medskip\\
\end{center}
{\it Tab.1: Some selected isotopes with reasonable half life limits for 
\bb. Shown are the present limits
for the half life $T_{1/2}^{0\nu}$, the resulting \ema - limit using the
matrix elements from \cite{ast}, as well
as some proposed limits of ongoing, upcoming or planned experiments on 
$T_{1/2}^{0\nu}$.}
\medskip\\
Clearly there are two strategies to follow for looking at the A-dependence:
The largest effect is expected 
if the two isotopes have the largest possible spread within A f.e. $\ca$ and $\nd$ or even $^{238}U$.
On the other hand much better experimental limits exist for the 
\obb half life for isotopes
like $\ger$ and $\teha$.
\begin{figure}[hhh]
\begin{center}
\epsfig{file=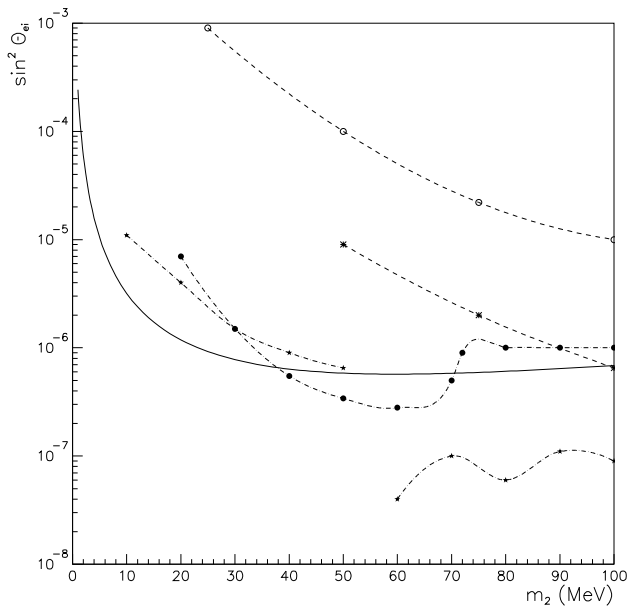,width=9cm,height=9cm}
\end{center}
\caption{\label{lego2} \it Comparison of the Ge-Te double beta decay 
limit
(solid line) on the mixing angle with other experiments. Shown are the 
results from [15] (open circles), [16] (asterix), [17] (stars)
and [18] (filled circles). It can be seen
that the double beta decay results give the best limit below about 35 
MeV. The region above the curves are excluded.}
\end{figure}
Fig.1a gives an idea of 
the variation in the mixing angle as a function of the heavy neutrino
mass for five different pairs of isotopes. On the ''small'' A side it is \ger (best half
life limit) and 
\ca (lowest A) on the ''higher'' side \teha (best half life limit) and \nd (largest A with reasonable
half life limit).
\begin{figure}[hhh]
\begin{center}
\begin{tabular}{cc}
\epsfig{file=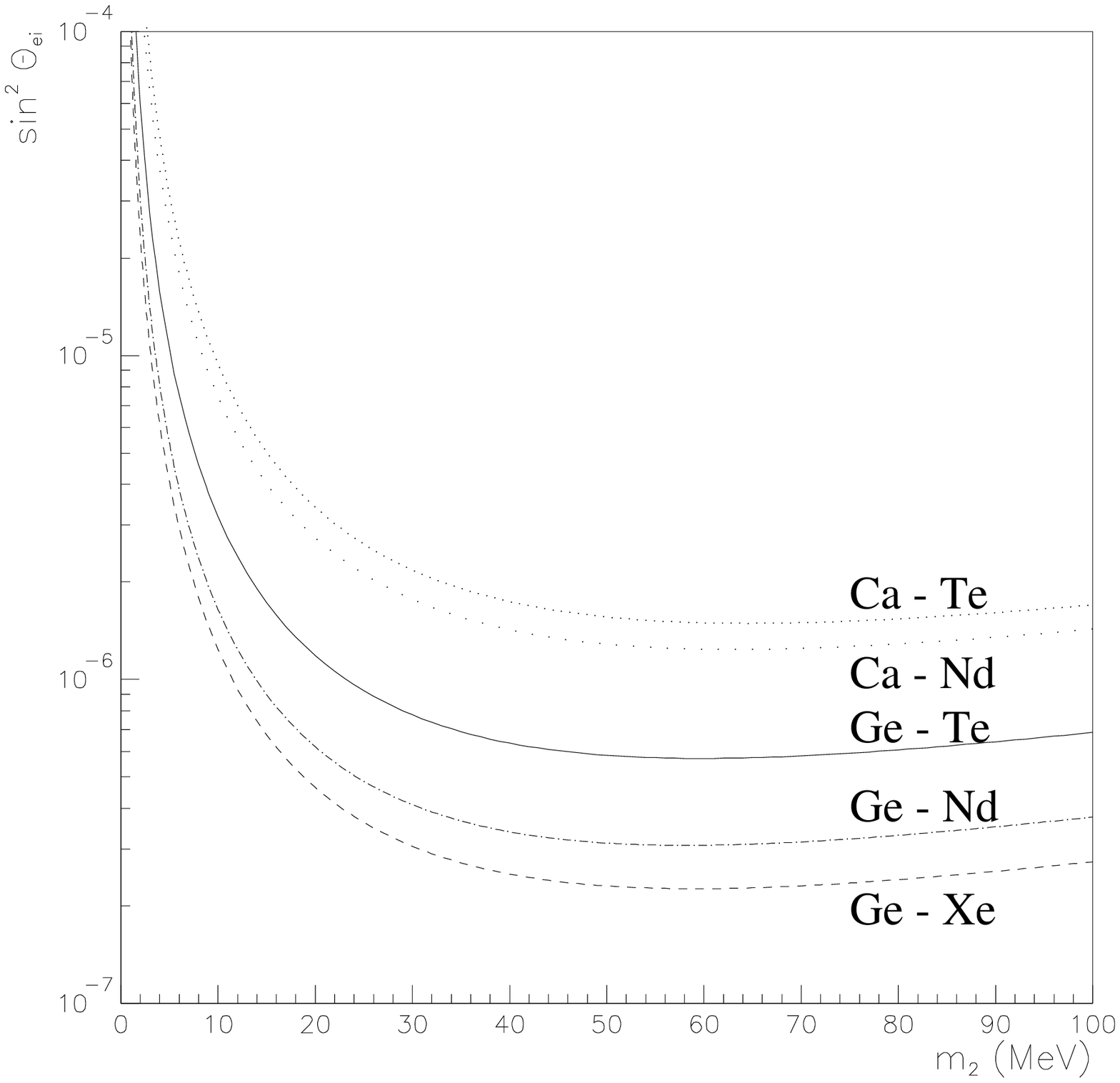,width=8cm,height=8cm}
&\epsfig{file=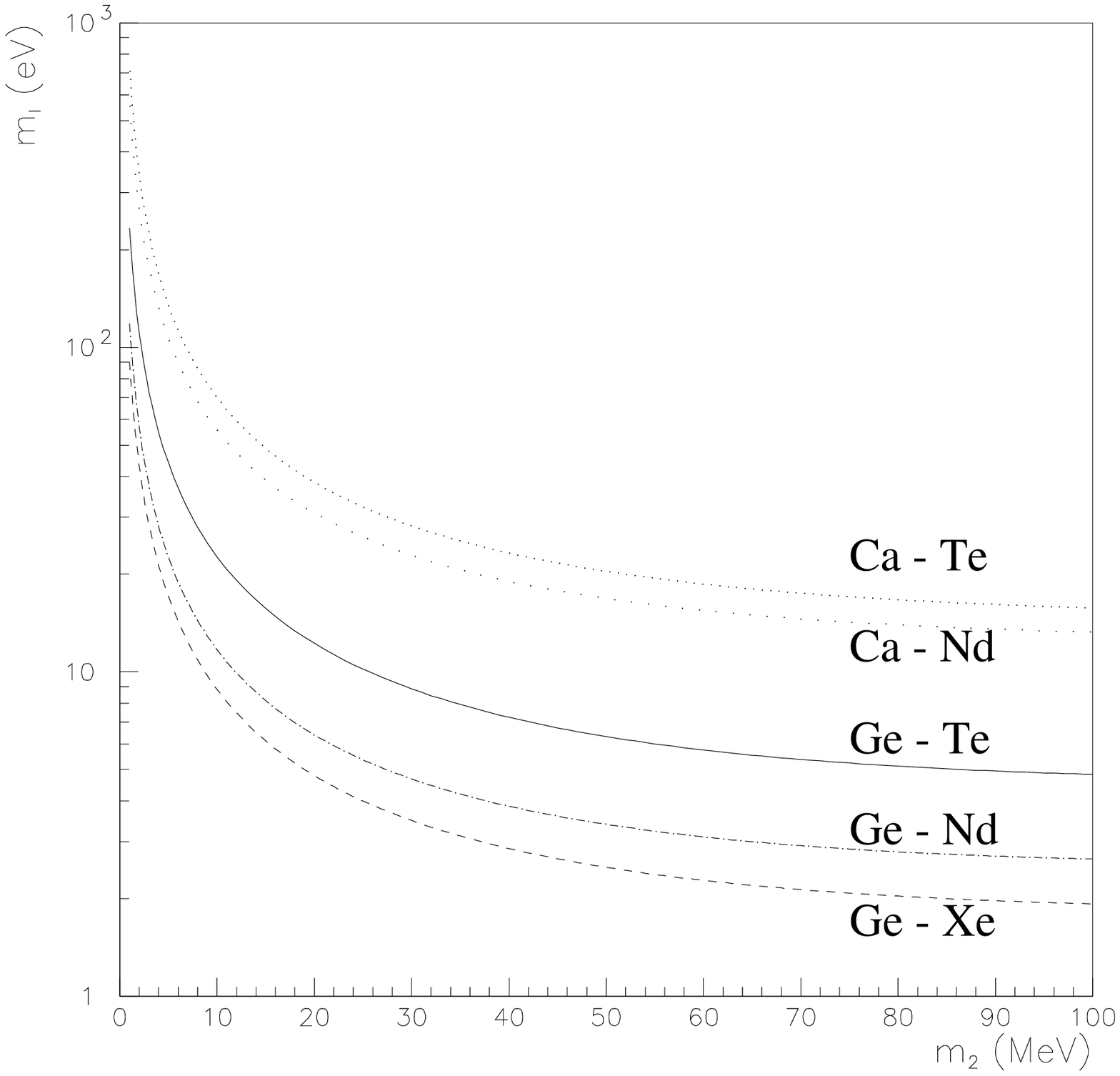,width=8cm,height=8cm}
\end{tabular}
\end{center}
\caption{\label{lego1} \it Same presentations as fig.1 for ongoing, 
upcoming or planned experiments and their proposed half lifes limits.
The best mixing angle
limit (left) is now given by the Ge-Xe pair, which also gives an upper
bound on $m_1$ of 2 eV assuming $m_2 \approx 100 MeV$.}
\end{figure}
It can be seen that the much
better experimental half life limits of $\ger$ and $\teha$ overcompensate the smaller
separation in A. Therefore this pair will be used for a comparison
with other experiments. Alternatively we can restrict the limit on the 
light neutrino
mass $m_1$ assuming a nearly perfect cancellation of $<m_{\nu_e}>$ 
in \ger by using a relation between two isotopes like \cite{grotz}
\be
m_1 < \frac{F(m_2,76)}{\mid F(m_2,76)-F(m_2,A) \mid} (<m_{\nu}(76)> + 
<m_{\nu}(A)>) \ee
where A is one of the other isotopes.
The value of the observable 
light \n mass as a function of the heavy one is shown in fig.1b.
Using the simple mixing scheme of eq.(\ref{fo11}) fig.2 shows the limits on 
the mixing parameter
$\mid U_{ei}\mid ^2 = sin^2 \theta_{ei}$ because of the A-dependence
in \bb in comparison with other experiments.
It can be seen that in the region below about 35 MeV this limit is the most 
stringent one. 
This can be improved by the ongoing,upcoming or planned future experiments.
Using the proposed half live limits of the experiments shown in Tab.1 limits
on the mass eigenstate and the mixing angle can be obtained as shown in 
fig.3.
In 
this case 
the \ger and \xehs data would dominate the bound, which is by a factor
of about 3 lower than the present limits. So far all the considerations are based on very general 
arguments.\\ Now look at
the standard model and assume that there are no extra neutrinos from
yet unknown physics.
The $Z^o$-resonance width results in 2.983 $\pm$ 0.025 
\cite{pdg}
neutrinos flavours lighter than 45 GeV.
Using the upper bounds on the different \n masses
as given in eq. (1-3) it is obvious that $m_2$ can only be associated 
with $\nu_{\tau}$.
Therefore the derived values for $\mid U_{ei}\mid ^2$ correspond to 
$\mid U_{e\tau}\mid ^2$ for $\Delta m^2 \approx 10^{12}$ eV$^2$. It 
should be mentioned that the    
bounds on $\mid U_{e\tau}\mid ^2$ are weakened by some orders of 
magnitude by including $\nu_{\mu}$ and assuming that there is a 
cancellation of the $\nu_{\mu}$ and $\nu_{\tau}$ contributions to \ema.\\
A direct consequence of the derived limit can be applied to the  
decay of $\nu_{\tau}
\ra \nu_e + e^+e^-$ because of the lifetime dependence of $\tau \propto \mid 
U_{ei}\mid 
^2$ (neglecting other decay channels, see \cite{mike}). Different limits 
on the lifetime
of $\nu_{\tau}$ as a function of its mass including the one obtained 
via \bb can be seen in fig. \ref{plane}.
By using only rather model independent limits most of the allowed parameter 
space can be ruled out.\\
\begin{figure}[hhh]
\begin{center}
\epsfig{file=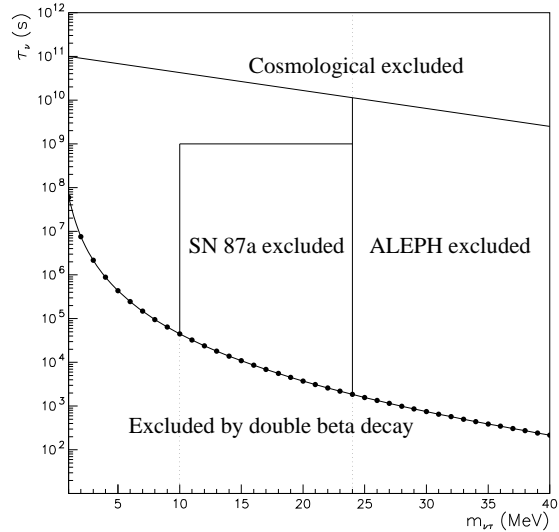,width=8cm,height=8cm}
\end{center}
\caption{\label{plane} \it Lifetime of $\nu_{\tau} \ra \nu_e + e^+ e^-$
against the tau-\n mass $m_{\nu_{\tau}}$. Shown is the exclusion due to
double beta decay (lower curve) 
The vertical line at 24 MeV corresponds to
the ALEPH-limit, the upper curve is 
excluded by cosmological arguments
(Overdensity of the universe). These two bounds are valid for any decay 
mode.
Also shown is the bound on the $\nu_{\tau} \ra \nu_{e,\mu} + \gamma  
(e^+ e^-)$ decay-mode given by SN 1987a [20].}
\end{figure}
As a 
second example consider the possible evidence for \n oscillations as seen
with the LSND-detector \cite{lsnd}. This would require a massive neutrino of $\approx$ 1 eV.
Assume that the mass of the electron \n is somewhere in the region of 1-4 
eV, which is still
allowed by beta decay experiments, a heavy \n between 1-24 MeV, and that
a perfect cancellation in \ger \obb occurs. This would imply values 
of \ema in other isotopes like \mo, \teha and \nd in the range
\bea
2.7 \cdot 10^{-3} eV < \ema < 0.25 eV  \quad (\mo)\\
5.5 \cdot 10^{-3} eV < \ema < 0.49 eV  \quad (\teha)\\
7.4 \cdot 10^{-3} eV < \ema < 0.63 eV  \quad (\nd) 
\eea
The upper bounds are in the region of upcoming or planned
experiments and such a scenario can be tested within the near future. 
\section*{Conclusions}
The effects of MeV \mas in \bb are investigated by comparing half life limits of
different isotopes. Using only experimental bounds the most stringent 
limits on the 
mixing matrix element $U_{ei}$ in the region from 1-35 MeV are obtained.
Using the bound on $U_{ei}$ a large part in the 
$\tau_{\nu_\tau}-m_{\nu_\tau}$ plane
for the decay mode $\nu_\tau \ra \nu_e e^+ e^-$ can be excluded. Whether 
there is
a Majorana $\nu_e$ somewhere in the eV region and a MeV $\nu_\tau$ 
can be tested or detected within the near future.
\section*{Acknowledgements}
I would like to thank P. Bamert, C. P. Burgess and R. N. Mohapatra for 
helpful comments.

\end{document}